\documentclass[preprint,aps,amsmath,amssymb,amsthm,nofootinbib]{revtex4}
\usepackage[mathscr]{euscript}
\usepackage{threeparttable}
\usepackage{subfigure}
\usepackage{bm}% bold math
\usepackage{textcomp}
\usepackage{graphicx}% Include figure files
\usepackage{subfigure}
\usepackage{overpic}
\usepackage{multirow}
\usepackage{floatrow}
\usepackage{float} %for fixing the figure
\usepackage{amsmath,amssymb,amsthm}
\usepackage{cases}
\usepackage[colorlinks=true,linkcolor=red]{hyperref}
\usepackage{xcolor}
\usepackage[normalem]{ulem}
\newfloatcommand{capbtabbox}{table}[][\FBwidth]

\begin{document}

\title{Coupled quintessence with a potential from supergravity exhibits sign-changing interaction}

\author{ Jincheng Wang\footnote{J.C.Wang@hunnu.edu.cn}}
	
\affiliation{Department of Physics,  Key Laboratory of Low Dimensional Quantum Structures
and Quantum Control of Ministry of Education, and Hunan Research Center of the Basic Discipline for Quantum Effects and Quantum Technologies, Hunan Normal University, Changsha, Hunan 410081, China}  

\author{ Hongwei Yu\footnote{Corresponding author: hwyu@hunnu.edu.cn}}
\affiliation{Department of Physics,  Key Laboratory of Low Dimensional Quantum Structures
and Quantum Control of Ministry of Education, and Hunan Research Center of the Basic Discipline for Quantum Effects and Quantum Technologies, Hunan Normal University, Changsha, Hunan 410081, China}

 \author{Puxun Wu\footnote{Corresponding author: pxwu@hunnu.edu.cn} }
\affiliation{Department of Physics,  Key Laboratory of Low Dimensional Quantum Structures
and Quantum Control of Ministry of Education, and Hunan Research Center of the Basic Discipline for Quantum Effects and Quantum Technologies, Hunan Normal University, Changsha, Hunan 410081, China}

\begin{abstract}
Quintessence with a potential motivated by supergravity (SUGRA) exhibits several intriguing features. Depending on its initial conditions, it can behave either as dynamical dark energy or effectively as a cosmological constant. Moreover, when quintessence is coupled to dark matter, the effective dark-energy equation of state can cross the phantom divide. In this paper, we test both coupled and uncoupled SUGRA quintessence models using DESI BAO, DES-Dovekie SNIa, and Planck CMB data. We find that current observations strongly favor a coupling between dark energy and dark matter, with the coupling parameter deviating from zero at more than $4\sigma$. The data also favor the branch of coupled SUGRA quintessence in which the energy transfer between the two dark sectors changes sign, leading to a crossing of the phantom divide by the effective dark-energy equation of state. Interestingly, this coupled SUGRA branch is statistically indistinguishable from dark energy described by the CPL parametrization, with only a very small difference in $\chi^2_\mathrm{min}$. Our results suggest that coupled quintessence with a SUGRA potential provides a field-theoretic realization of the evolving dark energy behavior favored by the latest observations.
\end{abstract}

	%\pacs{98.80.Cq, 04.50.Kd, 05.70.Fh}
	
	\maketitle
	%%%%%%%%%%%%%%%%%%%%%%%%%%%%%%%%%%%%%%%%%%%%%%%%%%
	
\section{Introduction}
\label{sec1}

Observations of type Ia supernovae (SNIa) first established that the Universe is undergoing accelerated expansion~\cite{riess1998observational,perlmutter1999measurements}. This picture has since been reinforced by baryon acoustic oscillation (BAO) and cosmic microwave background (CMB) measurements~\cite{eisenstein2005detection,planck2018cosmo}. Within general relativity, the accelerated expansion is usually attributed to a dark-energy component with sufficiently negative pressure. The simplest candidate is the cosmological constant $\Lambda$, whose equation of state is fixed at $w=-1$. However, dark energy may also be dynamical, with an equation of state that evolves over cosmic time. A natural realization of dynamical dark energy is quintessence, described by a canonical scalar field slowly rolling along its potential and satisfying $-1\leq w\leq 1$~\cite{Caldwell1998,ratra1988cosmological,Zlatev1998}. By contrast, a scalar field with $w<-1$ is usually referred to as phantom dark energy~\cite{Caldwell2002}. Such models typically require a negative kinetic term, which leads to quantum instabilities~\cite{Carroll2003,cline2004phantom}.

Combining a cosmological constant with cold dark matter gives the $\Lambda$CDM model, which has been remarkably successful in explaining a wide range of cosmological observations and is therefore regarded as the standard model of cosmology. Recent BAO measurements from the Dark Energy Spectroscopic Instrument (DESI), however, suggest that the late-time evolution of the Universe may deviate from the prediction of $\Lambda$CDM~\cite{Adame2024desi,DESIDR2}. This has strengthened the possibility that dark energy is not strictly constant but dynamical. Using the widely adopted Chevallier--Polarski--Linder (CPL) parametrization,
 $w(z)=w_0+w_a \frac{z}{1+z}$, where $w_0$ and $w_a$ are constants and $z$ is the 
redshift~\cite{chevallier2001accelerating,linder2003exploring}, 
the DESI Collaboration found that the combination of DESI BAO and Planck CMB data favors the spatially flat $w_0w_a$CDM model over $\Lambda$CDM at about $2.5\sigma$~\cite{Adame2024desi}. This preference increases to approximately $3.9\sigma$ when the DES-SN5YR SNIa data are included~\cite{Adame2024desi,dessn5yr2024}. Moreover, the data favor an evolving equation of state that crosses the phantom divide, transitioning from $w<-1$ to $w>-1$ in the recent past.

Such a crossing of the phantom divide cannot be achieved by a single minimally coupled quintessence field without introducing phantom degrees of freedom. To avoid the instabilities associated with phantom fields while still realizing phantom-divide crossing, one may extend the simple quintessence scenario by allowing energy exchange between dark energy and dark matter. In this case, the effective equation of state of quintessence can cross $w=-1$ even though the underlying scalar field remains canonical. This possibility has renewed interest in coupled scalar-field dark-energy models in the DESI era~\cite{wang2024further,chakraborty2024hint,giare2024desiide,li2024desiide,chakraborty2025desi,shah2025interacting,silva2025newconstraints,datadriven2025coupled,pan2025afterdesi,li2025cosmicsign,guedezounme2025phantom,petri2025darkdegeneracy,yang2025variable,li2025updatedide,li2026strongevidence,figueruelo2026late,gomezvalent2026desi,sahoo2025momentum,pourtsidou2025exp,samanta2025exploring,beltran2026ceq,antusch2026guide,artola2026cplinteraction,wang2026planckdesi}.

Although the cosmological dynamics of coupled quintessence depends on the coupling strength, it is primarily governed by the scalar-field potential. A commonly studied potential is the inverse power-law form, $V(\phi)\sim \phi^{-\alpha}, $
originally proposed by Ratra and Peebles~\cite{ratra1988cosmological}. A key feature of this potential is that it admits tracker evolution for the scalar field~\cite{wang2026planckdesi,pettorino2012constraints,pettorino2013testing,planck2015de,gomezvalent2020h0}. Exponential potentials, $ V(\phi)\sim e^{-\lambda\phi}, $
constitute another important class, as they allow well-known scaling or stationary solutions~\cite{amendola2000coupled,wetterich1988cosmology,copeland1998exponential,mifsud2019siren}. Recent studies have also shown that quintessence with an exponential potential can remain frozen at early times and begin to evolve only at late times~\cite{ramadan2024desi,desouza2025desi,wang2026signswitch}.

A realistic model of a quintessential tracking field is expected to have a possible origin in high-energy physics. One such framework is supergravity (SUGRA). Within this framework, Brax and Martin~\cite{brax1999quintessence} derived the SUGRA scalar potential: $
V(\phi)\sim \phi^{-\alpha}e^{\phi^2/2}. $
For sufficiently small field values, this potential reduces to the inverse power-law form and therefore admits tracker behavior. At the same time, the SUGRA potential possesses a global minimum at $\phi\sim\sqrt{\alpha}$. Consequently, in the minimally coupled case, SUGRA quintessence can behave effectively as a cosmological constant when the scalar field settles near this minimum.

These properties motivate two distinct classes of initial conditions. In the first scenario, the field starts on the inverse power-law tail, undergoes tracker evolution, and rolls toward the minimum of the potential. In the second scenario, the field is initially placed near the minimum, so that in the absence of coupling between dark energy and dark matter the model reduces effectively to $\Lambda$CDM. If such a coupling is present, however, the interaction displaces the scalar field from the minimum, after which the field eventually rolls back toward it. This bounce-like behavior naturally induces a sign change in the energy transfer between the two dark sectors and can drive the effective dark-energy equation of state across the phantom divide.

Quintessence with a SUGRA potential, both with and without coupling to dark matter, has been extensively studied using WMAP-era data~\cite{mainini2005singlefield,bonometto2006cdecmb,lavacca2008highernu,lavacca2009wmap,kristiansen2010neutrino,baldi2011bouncing,baldi2012codecs,carbone2013lensing}. Whether this model remains compatible with, or is favored by, the latest DESI BAO measurements remains an open question. Motivated by this issue, we test SUGRA quintessence using DESI DR2 BAO data in combination with DES-Dovekie SNIa and Planck CMB observations. We compare both coupled and uncoupled SUGRA quintessence models with coupled quintessence based on the inverse power-law potential and with the phenomenological $w_0w_a$CDM model. Our goal is to determine whether coupled SUGRA quintessence can provide a field-theoretic realization of the evolving dark-energy behavior preferred by recent observations. As we will show, one branch of coupled SUGRA quintessence achieves a statistical fit close to that of the $w_0w_a$CDM model, making it a compelling physically motivated alternative to a purely phenomenological description of evolving dark energy.
%%%%%%%%%%%%%%%%%%%%%%%

\section{Model and Data}
\label{sec2}

\subsection{Quintessence with a SUGRA potential}

We consider a cosmological model in which quintessence scalar field $\phi$  couples with  CDM,  
while baryon and radiation remain minimally coupled.  Thus, the  action of the system can be expressed as follows:
\begin{equation}
\begin{aligned}
S ={}& \int d^4x \sqrt{-g}\left[\frac{M_{\rm Pl}^2}{2}R-\frac{1}{2}g^{\mu\nu}\partial_{\mu}\phi\partial_{\nu}\phi-V(\phi)\right] \\
& +S_c\!\left[e^{-2\beta \phi/M_{\rm Pl}}g_{\mu\nu},\psi_c\right]  +S_b\!\left[g_{\mu\nu},\psi_b\right]
 +S_r\!\left[g_{\mu\nu},\psi_r\right],
\end{aligned}
\label{action_cq}
\end{equation}
where $M_{\rm Pl}\equiv 1/\sqrt{8\pi G}$ is the reduced Planckian  constant, $g_{\mu\nu}$ is the metric tensor, $R$ is the Ricci scalar,  $V(\phi)$ is the potential of quintessence, and dimensionless constant $\beta$ describes the coupling strengthen between two  dark sectors.  $\psi_c$, $\psi_b$, and $\psi_r$ represent CDM, baryon and radiation, respectively. When $\beta=0$, this model reduces to the minimally coupling case.   

Varying Eq.~(\ref{action_cq}) yields the covariant energy exchange equations
\begin{equation}
\nabla_{\mu}T^{\mu\nu}_{(c)}=-\frac{\beta}{M_{\rm Pl}}\rho_c\nabla^{\nu}\phi,
\qquad
\nabla_{\mu}T^{\mu\nu}_{(\phi)}=+\frac{\beta}{M_{\rm Pl}}\rho_c\nabla^{\nu}\phi,
\label{covariant_exchange}
\end{equation}
with the baryon and radiation sectors separately conserved. In a spatially flat FLRW background,  Eq.~(\ref{covariant_exchange}) reduces to
\begin{eqnarray}
&&\dot{\rho}_c+3H\rho_c=-Q, \label{Eq3}
\\
&& \dot{\rho}_{\phi}+3H(\rho_{\phi}+p_{\phi})=+Q.
\label{Eq4}
\end{eqnarray}
Here $\rho_c$ and $\rho_\phi$ are respectively the energy densities of CDM and dark energy,   $p_\phi$ is the pressure of dark energy,  $H\equiv \dot a/a$ with $a$ being the scale factor and an overdot denoting a derivative with respect to time, and the interaction term $Q$ has the form
\begin{eqnarray}\label{Eq5}
Q=\frac{\beta}{M_{\rm Pl}}\rho_c\dot{\phi}.
\end{eqnarray}
$Q>0$ denotes energy transfer from CDM to dark energy, while 
$Q<0$ signifies energy transfer from dark energy to CDM. The baryon and radiation sectors are separately conserved, and their energy densities maintain the standard scaling behaviors, given by $\rho_b\propto a^{-3}$ and $\rho_r\propto a^{-4}$.

Substituting Eq.~(\ref{Eq5}) into Eq.~(\ref{Eq3}), one can obtain
\begin{equation}
\rho_c=\rho_{c0}a^{-3}
\exp\!\left[-\frac{\beta}{M_{\rm Pl}}(\phi-\phi_0)\right],
\label{Eq6}
\end{equation}
where the subscript  $0$ represents the present value.  Using $\rho_{\phi}=\frac{1}{2}\dot{\phi}^{\,2}+V(\phi)$ and $p_{\phi}=\frac{1}{2}\dot{\phi}^{\,2}-V(\phi)$,  one can derive the dynamical equation of the scalar field  from Eq.~(\ref{Eq4})
\begin{equation}
\ddot{\phi}+3H\dot{\phi}+V_{\mathrm{eff},\phi}=0
\label{Eq7}
\end{equation}
with 
\begin{equation}
V_{{\rm eff},\phi}=V_{,\phi}-\frac{\beta}{M_{\rm Pl}}\rho_c .
\label{Eq8}
\end{equation}
Here $V_{,\phi}=dV/d\phi$. The second term on the rhs of Eq.~(\ref{Eq8}) can be interpreted as an effective force resulting from the coupling with CDM. Eqs.~(\ref{Eq6}) and (\ref{Eq7}) explicitly demonstrate  that the coupling affects the inferred evolution of the dark sectors. 

Assuming that the coupled dark sector energy can be divided into CDM and an effective dark energy, we find that the energy density of this effective dark energy can be expressed as follows:
\begin{equation}
\rho_{{\rm DE},{\rm eff}}=\rho_\phi+\rho_c-\rho_{c0}a^{-3}.
\end{equation}
Then, the equation of state for this effective dark energy takes the form
\begin{equation}
w_{{\rm DE},{\rm eff}}=\frac{w_\phi}{1+ (\rho_c-\rho_{c0}a^{-3})/\rho_\phi}.
\label{Eq_wDEeff}
\end{equation}
Here $w_\phi=p_\phi/\rho_\phi$. 

\begin{figure}[t]
\centering
\includegraphics[width=0.72\textwidth]{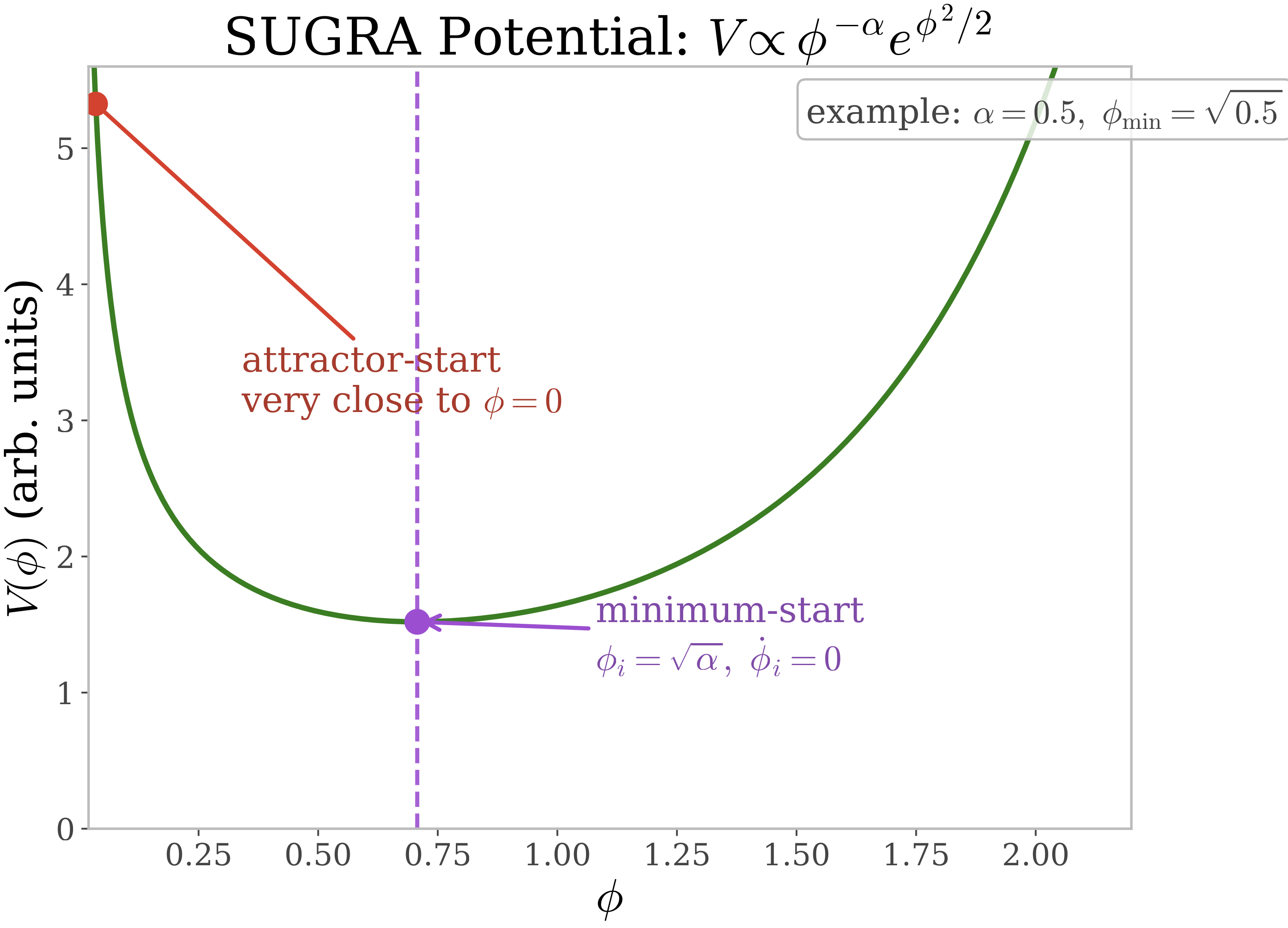}
\caption{The SUGRA potential and  two SUGRA initialization branches (attractor-start and minimum-start ). }
\label{fig1}
\end{figure}

We now specify the bare potential of the quintessence scalar field, focusing on  the SUGRA potential motivated by supergravity theory~\cite{brax1999quintessence},
\begin{equation}
V(\phi)=V_0\left(\frac{\phi}{M_{\rm Pl}}\right)^{-\alpha}
\exp\!\left(\frac{\phi^2}{2M_{\rm Pl}^2}\right).
\label{V_SUGRA}
\end{equation}
Here, $V_0$ sets the overall dark energy scale and is determined by matching the current dark energy density, while $\alpha$, a positive constant, controls the shape of the potential. For $\phi\ll M_{\rm Pl}$, the exponential part becomes negligible and the model exhibits the Ratra-Peebles inverse power law behavior~\cite{ratra1988cosmological}. As the field value increases, the exponential part becomes significant, causing the potential to rise. Thus, the SUGRA potential has a global minimum, which occurs  at  
\begin{equation}
\phi_{\rm min}=\sqrt{\alpha}\,M_{\rm Pl}.
\label{phi_min_sugra}
\end{equation}
These features can be observed  in Fig.~\ref{fig1}, where the SUGRA potential is plotted. 

 Based on the features of the SUGRA potential,   there are two natural initialization choices for the quintessence field. First, the field may begin on the inverse power law tail and thus initially follow the usual tracker type evolution; we refer to this as {\it the attractor-start branch}. Second, the field may be initialized at $\phi_{\rm min}$ with negligible velocity and then  displaced by the CDM coupling through the effective force described in Eq.~(\ref{Eq8}); we call this {\it the minimum-start branch}. In Refs.~\cite{baldi2011bouncing,baldi2012codecs}, the minimum-start branch   is also referred to  the bouncing coupled dark energy scenarios.  These two different branches have been shown clearly in Fig.~\ref{fig1}.   

\subsection{Linear perturbations}

When using CMB observations to constrain quintessence with a SUGRA potential, it is essential to account for  the evolution of perturbed energy components. The interaction within dark sectors affects the dynamics of   these perturbations, necessitating the derivation of their evolution equations.      Following the standard CAMB convention and previous coupled quintessence treatments~\cite{amendola2000coupled,pettorino2008cqe,ma1995cosmological}, we will adopt  the synchronous gauge. Consequently, the perturbed metric can be expressed as 
\begin{equation}
ds^2=a^2(\tau)\left[-d\tau^2+\left(\delta_{ij}+h_{ij}\right)dx^idx^j\right],
\label{sync_metric}
\end{equation}
where $\tau$ is conformal time.  Then, we find that the scalar field perturbation $\delta\phi$ obeys the following equation 
\begin{equation}
\delta\phi''+2\mathcal H\delta\phi'+\left(k^2+a^2V_{,\phi\phi}\right)\delta\phi+\frac{h'}{2}\phi'
=a^2\frac{\beta}{M_{\rm Pl}}\rho_c\delta_c,
\label{dphi_eq}
\end{equation}
where a prime denotes differentiation with respect to $\tau$, $k$ is the comoving wavenumber, $\mathcal H\equiv a'/a$, $h=\delta^{ij}h_{ij}$ and $\delta_c\equiv \delta\rho_c/\rho_c$ with  $\delta\rho_c$ representing the perturbation of the CDM energy  density. 
We obtain that $\delta_c$ is governed by the equation 
\begin{equation}
\delta_c'=-\theta_c-\frac{h'}{2}-\frac{\beta}{M_{\rm Pl}}\delta\phi',
\label{delta_c_eq}
\end{equation}
where  $\theta_c$ denotes the CDM velocity divergence perturbation, which satisfies the equation  
\begin{equation}
\theta_c'=-\mathcal H\theta_c+\frac{\beta}{M_{\rm Pl}}\left(\phi'\theta_c-k^2\delta\phi\right).
\label{theta_c_eq}
\end{equation}
We modify the CAMB~\cite{lewis2000camb,li2014ppf,hu2014eftcamb,li2023idecamb} to compute the evolution of perturbations.

\subsection{Datasets}

The observational datasets employed in this work include:

\begin{itemize}

\item \textbf{DESI BAO DR2}: The BAO data used in analysis  comes from    the second data release of the DESI experiment~\cite{DESIDR2,DESIDR2b}. This dataset combines information from various sources, including bright galaxies, luminous red galaxies, emission line galaxies, quasars, and Lyman-$\alpha$ forest tracers,  providing  precise distance measurements over a broad redshift range.

\item \textbf{DES-Dovekie supernovae}: We utilize the recalibrated DES-Dovekie SNIa compilation~\cite{desdovekie2025}, which updates the DES five year supernova analysis~\cite{dessn5yr2024} through improved photometric cross calibration. This compilation offers   a  late time distance probe for studies of dark energy.

\item \textbf{Planck CMB}: For the CMB data, we employ the Planck 2018 low $\ell$ TT and EE likelihoods for the large scale temperature and polarization information~\cite{planck2018cosmo}. Additionally, we incorporate the CamSpec high $\ell$ TTTEEE likelihood, which is based on the final Planck PR4/NPIPE release~\cite{rosenberg2022camspec}. Furthermore, we include the Planck PR4 CMB lensing likelihood, which is reconstructed from the NPIPE maps~\cite{carron2022pr4lensing}.

\end{itemize}

\section{Results and discussions}
\label{sec3}

We now present the observational constraints on minimally and non-minimally coupled quintessence with a SUGRA potential (the coupled case is abbreviated as: CQ-SUGRA). We will consider two  physically distinct branches: attractor-start and minimum-start, as shown in Fig.~\ref{fig1}. For comparisons, we also analyze $\Lambda$CDM, $w_0w_a$CDM,  and coupled quintessence with an inverse power law potential (CQ-PL). The uniform prior ranges adopted for the cosmological and model parameters are summarized in Table~\ref{tab1}. We use $\Delta\chi^2$, which is defined relative to the best fit $\Lambda$CDM model, to compare different models. 

\begin{table}[t]
\centering
\caption{Uniform priors for free parameters used in the MCMC.}
\label{tab1}
\renewcommand{\arraystretch}{1.25}

\newcommand{\wA}{1.75cm}
\newcommand{\wB}{2.75cm}
\newcommand{\wC}{1.0cm}
\newcommand{\wD}{2.5cm}

\begin{tabular}{l c|l c}
\hline\hline
\multicolumn{2}{c|}{Cosmological Parameters} & \multicolumn{2}{c}{Model Parameters} \\
\hline
\makebox[\wA][l]{$\Omega_b h^2$} & \makebox[\wB][c]{$U[0.005,\,0.1]$}
& \makebox[\wC][c]{$w_0$} & \makebox[\wD][c]{$U[-3,\,1]$} \\

\makebox[\wA][l]{$\Omega_c h^2$} & \makebox[\wB][c]{$U[0.001,\,0.99]$}
& \makebox[\wC][c]{$w_a$} & \makebox[\wD][c]{$U[-3,\,2]$} \\

\makebox[\wA][l]{$H_0$} & \makebox[\wB][c]{$U[20,\,100]$}
& \makebox[\wC][c]{$\alpha$} & \makebox[\wD][c]{$U[0,\,20]$} \\

\makebox[\wA][l]{$\ln(10^{10} A_s)$} & \makebox[\wB][c]{$U[1.61,\,3.91]$}
& \makebox[\wC][c]{$\beta$} & \makebox[\wD][c]{$U[-0.5,\,0.5]$} \\

\makebox[\wA][l]{$n_s$} & \makebox[\wB][c]{$U[0.8,\,1.2]$}
& & \\

\makebox[\wA][l]{$\tau$} & \makebox[\wB][c]{$U[0.01,\,0.8]$}
& & \\
\hline\hline
\end{tabular}
\end{table}

\subsection{results}

\begin{table*}[t]
\centering
\caption{Summary of the mean observational constraints  with $1\sigma$ CL for $\Lambda$CDM,   SUGRA quintessence, and  CQ-SUGRA. Here $\Delta \chi^2$ is defined relative to the best fit $\Lambda$CDM.}
\label{tab2}
\small
\setlength{\tabcolsep}{5pt}
\resizebox{\textwidth}{!}{%
\begin{tabular}{lccccc}
\hline\hline
Parameter & $\Lambda$CDM &
\begin{tabular}{c}SUGRA\\attractor-start\end{tabular} & \begin{tabular}{c}CQ-SUGRA\\attractor-start\end{tabular} & \begin{tabular}{c}CQ-SUGRA\\minimum-start, $\beta>0$\end{tabular} & \begin{tabular}{c}CQ-SUGRA\\minimum-start, $\beta<0$\end{tabular} \\
\hline
$\alpha$ & $-$ & $0.066^{+0.017}_{-0.054}$ & $0.86^{+0.47}_{-0.86}$ & unconstrained & $10.90^{+9.10}_{-6.33}$ \\
$\beta$ & $-$ & $-$ & $0.049^{+0.013}_{-0.009}$ & $0.049^{+0.010}_{-0.007}$ & $-0.045^{+0.006}_{-0.009}$ \\
$H_0$ & $68.09\pm 0.27$ & $67.73^{+0.37}_{-0.32}$ & $68.14\pm 0.54$ & $67.70\pm 0.28$ & $67.43\pm 0.35$ \\
%\hline\hline
$\Omega_m$ & $0.3037\pm 0.0035$ & $0.3062\pm 0.0038$ & $0.2973\pm 0.0049$ & $0.3076\pm 0.0036$ & $0.3100\pm 0.0041$ \\
%$S_8$ & $0.8107\pm 0.0078$ & $0.8091\pm 0.0079$ & $0.8183\pm 0.0089$ & $0.8270\pm 0.0088$ & $0.8239\pm 0.0088$ \\
\hline\hline
$\Delta \chi^2$ & $0$ & $-0.91$ & $-8.38$ & $-12.90$ & $-13.09$ \\
\hline\hline
\end{tabular}
}
\end{table*}

\subsubsection{Attractor-start branch}
\begin{figure*}[t]
\centering
\begin{minipage}{0.3\textwidth}
\centering
\includegraphics[width=\linewidth]{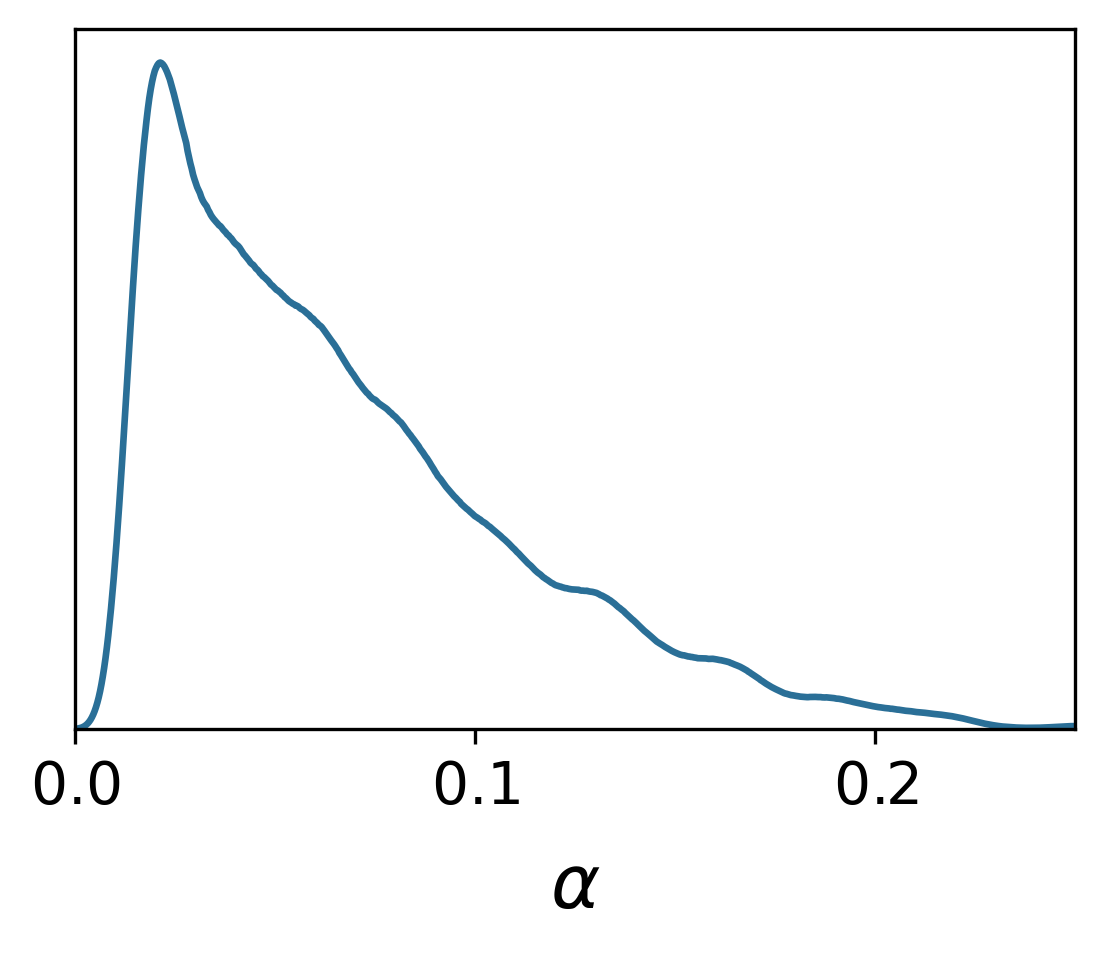}
\end{minipage}
\hspace{0.06\textwidth}
\begin{minipage}{0.34\textwidth}
\centering
\includegraphics[width=\linewidth]{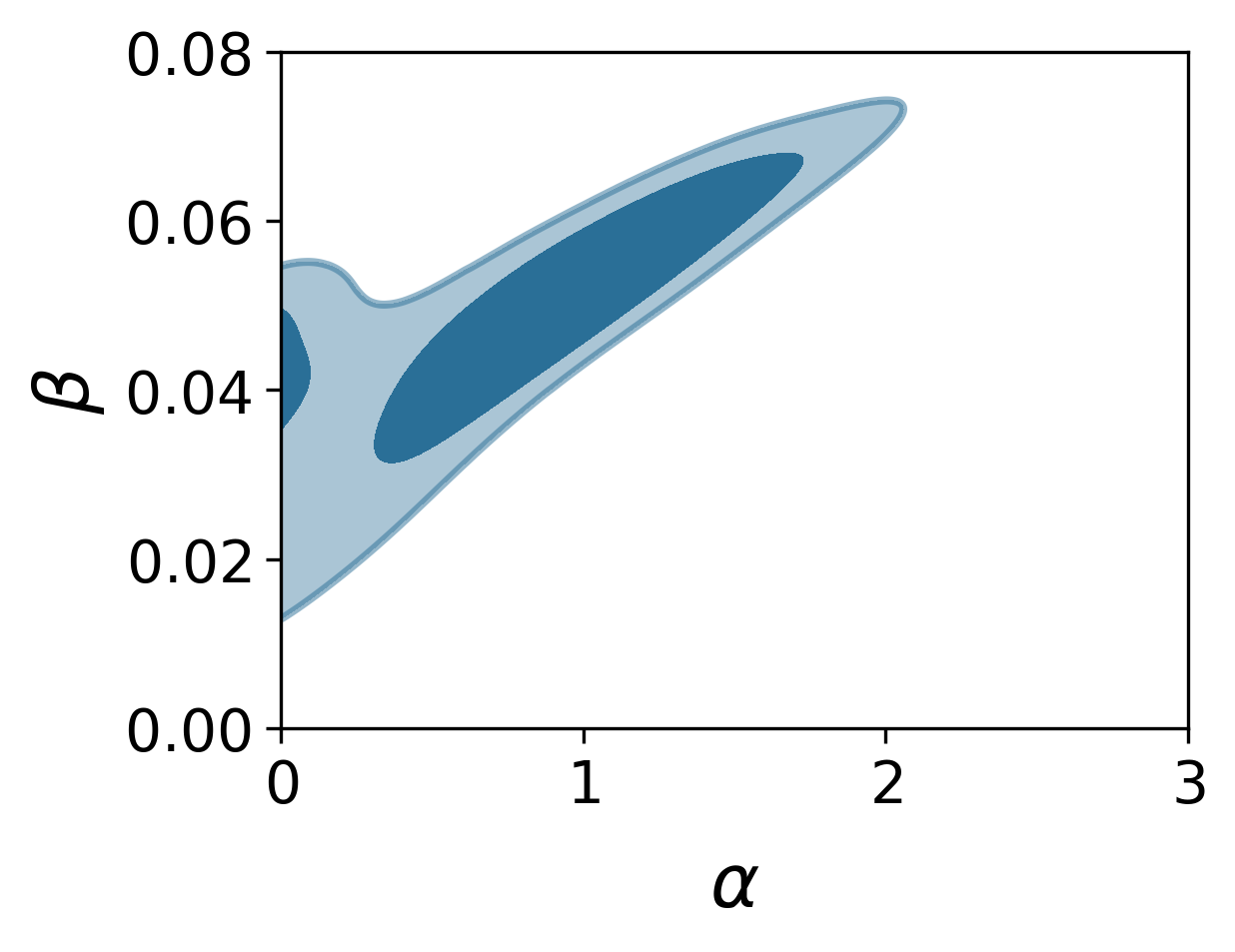}
\end{minipage}
\caption{Posterior distributions for the SUGRA potential parameters. Left: the posterior of $\alpha$ in the case of minimally coupled SUGRA. Right: the joint posterior of $(\alpha,\beta)$ in CQ-SUGRA.}
\label{fig2}
\end{figure*}

The attractor-start branch initializes the field on the inverse power law tail of the SUGRA potential. For the minimally coupling case, the constraints on free parameters are summarized in Tab.~\ref{tab2}, and the posterior distribution of $\alpha$  is plotted in the left panel of Fig.~\ref{fig2}. We  find that  $\alpha=0.066^{+0.017}_{-0.054}$, $\Omega_{m}=0.3062\pm0.038$ and $H_0=67.73^{+0.37}_{-0.32}$ km/s/Mpc at $1\sigma$ confidence level (CL). The allowed values of $\Omega_m$ and $H_0$ align well with those obtained in the $\Lambda$CDM model ($\Omega_{m}=0.3037\pm0.035$ and $H_0=68.09 \pm0.27$ km/s/Mpc). Moreover, the  result for $H_0$ is  compatible with the Planck CMB measurement, $H_0=67.4\pm0.5$ km/s/Mpc~\cite{planck2018cosmo}, but significantly smaller than the late-time distance ladder determination, $H_0=73.04\pm 1.04$ km/s/Mpc~\cite{riess2022shoes}.  The value of $\Delta \chi^2$ is $-0.91$, indicating that quintessence with a SUGRA potential is no significantly   favored  over the cosmological constant dark energy, despite having  an additional  free parameter.

For the coupled case, we present the posterior distributions of $\alpha$ and $\beta$ in the  right panel of Fig.~\ref{fig2} and summarize the allowed regions of model parameters in Tab.~\ref{tab2}. We find that the value of $\Omega_m$ is slightly smaller than that of the minimally coupling case, while the value of $H_0$ is slightly larger, although it remains significantly lower than $H_0=73.04\pm 1.04$ km/s/Mpc from the late-time distance observations~\cite{riess2022shoes}.  We observe   that $\alpha=0.86^{+0.47}_{-0.86}$ at $1\sigma$ CL with its uncertainty being much smaller than that in the minimally coupled case.  The data prefer a positive coupling, $\beta=0.049^{+0.013}_{-0.009}$, which  deviates from zero at more than $4\sigma$. This strongly indicates that the coupling between quintessence and CDM is favored  by observations.

\begin{figure*}[t]
\centering
\begin{minipage}{0.48\textwidth}
\centering
\includegraphics[width=\linewidth]{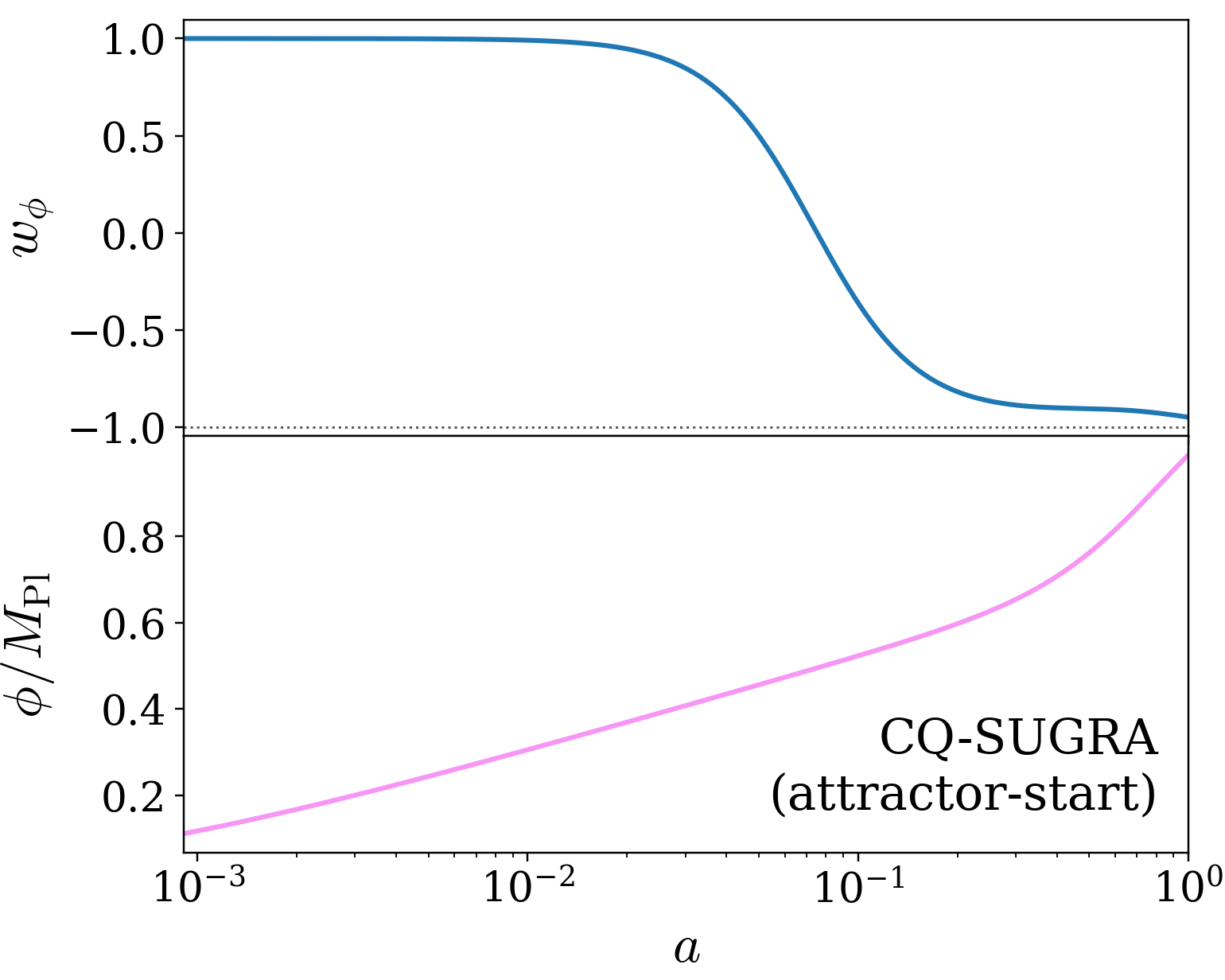}
\end{minipage}\hfill
\begin{minipage}{0.48\textwidth}
\centering
\includegraphics[width=\linewidth]{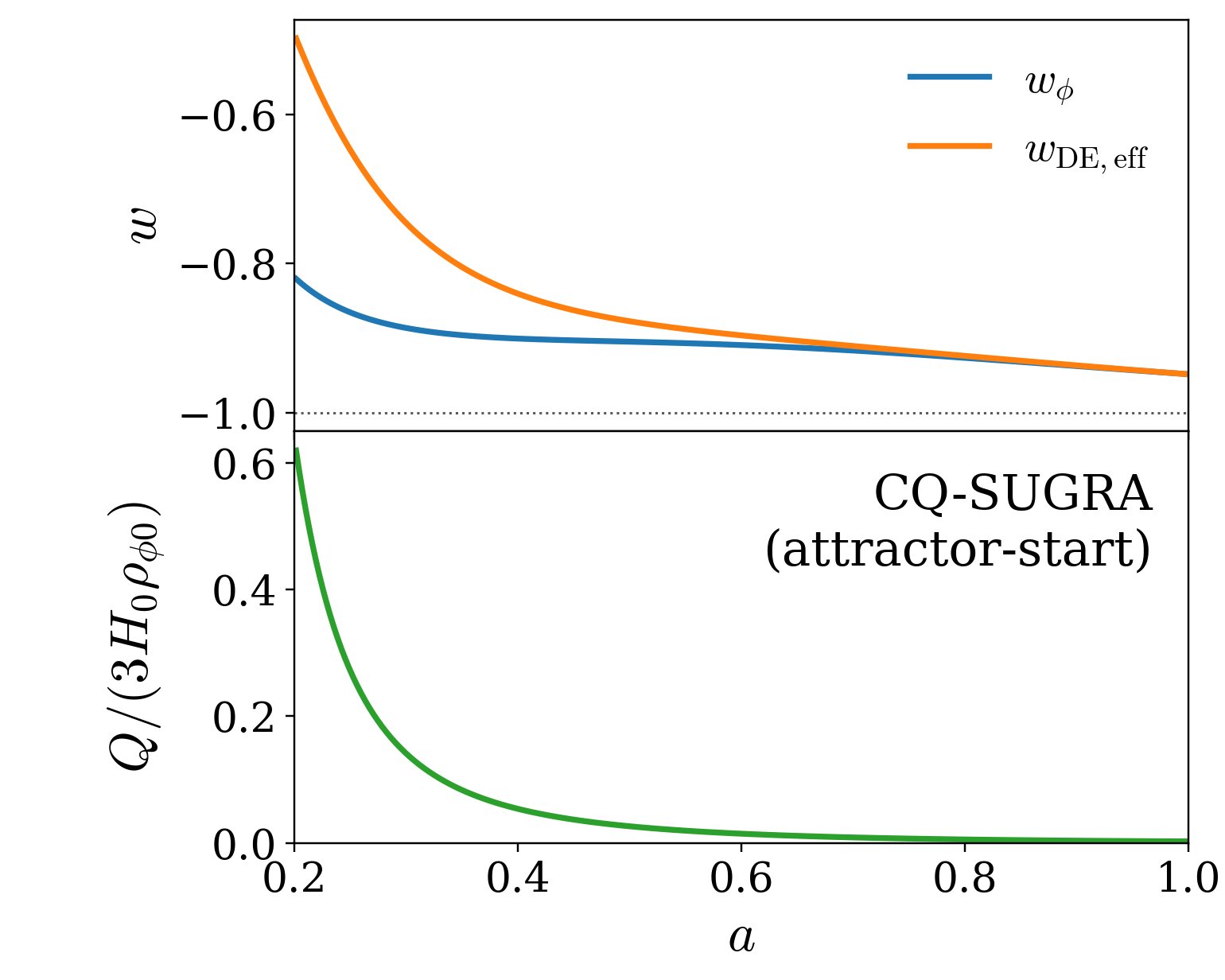}
\end{minipage}
\caption{Evolutions of  $w_\phi$,  $w_\mathrm{DE,eff}$,  $\phi$, and  $Q$ for the attractor-start CQ-SUGRA branch.}
\label{fig3}
\end{figure*}

Fig.~\ref{fig3} shows the evolutions of $\phi$, $w_\phi$, $w_{{\rm DE},{\rm eff}}$, and $Q$. From this,  we observe that the scalar field rolls down along its potential with cosmic expansion,  as its value consistently  increases with the scale factor $a$, indicating  that $\dot{\phi}>0$ is always satisfied. Therefore, we have $Q=\frac{\beta}{M_\mathrm{Pl}}\rho_c \dot{\phi}>0$ since a positive $\beta$ is supported by observations. This suggests  that  energy transfer occurs from CDM to the scalar field, a result that is clearly confirmed  in the right panel of Fig.~\ref{fig3}, where the evolution of $Q$ is plotted and remains positive. Moreover, $Q$ decreases with cosmic expansion, indicating that  the coupling between dark energy and CDM becomes weaker over time. In the high redshift region, $w_\phi$ is close to 1, suggesting that the scalar field is dominated by its kinetic energy in this phase. As cosmic expansion continues, the scalar field transitions into an era dominated by  potential energy,  leading to an accelerating expansion of the universe. $w_{{\rm DE},{\rm eff}}$ remains above $-1$ and approaches $w_\phi$ as the universe evolves toward the present epoch.  Therefore, the effective equation of state for dark energy in the attractor-start CQ-SUGRA branch does not cross  the phantom divide line.

  The difference in $\chi^2_\mathrm{min}$ between CQ-SUGRA at the attractor-start   branch and   $\Lambda$CDM is $\Delta\chi^2=-8.38$, indicating  that the former is more consistent with observations.

\subsubsection{Minimum-start branch}

At this branch,  the field is initially positioned  at the minimum of the SUGRA potential. If there is no coupling between quintessence and CDM, the field always stays at this minimum and quintessence behaves like a cosmological constant. The corresponding cosmological model reduces to  the $\Lambda$CDM.  However,  when  a coupling between quintessence and CDM is introduced, this coupling exerts an effective force on the motion equation of the scalar field,  as shown in Eq.~(\ref{Eq8}),  causing the field to be  displaced from its minimum.  As a result, this  branch possesses initial conditions and evolutions of the scalar field that differ from those in the attractor-start case. We find the value of $\chi^2$ has two different local minima: one for $\beta>0$ and the other for $\beta<0$.  This motivates us to  separate  our discussions into these two cases.

\begin{itemize}
\item $\beta>0$:
\end{itemize}

\begin{figure}[t]
\centering
\begin{minipage}{0.36\textwidth}
\centering
\includegraphics[width=\linewidth]{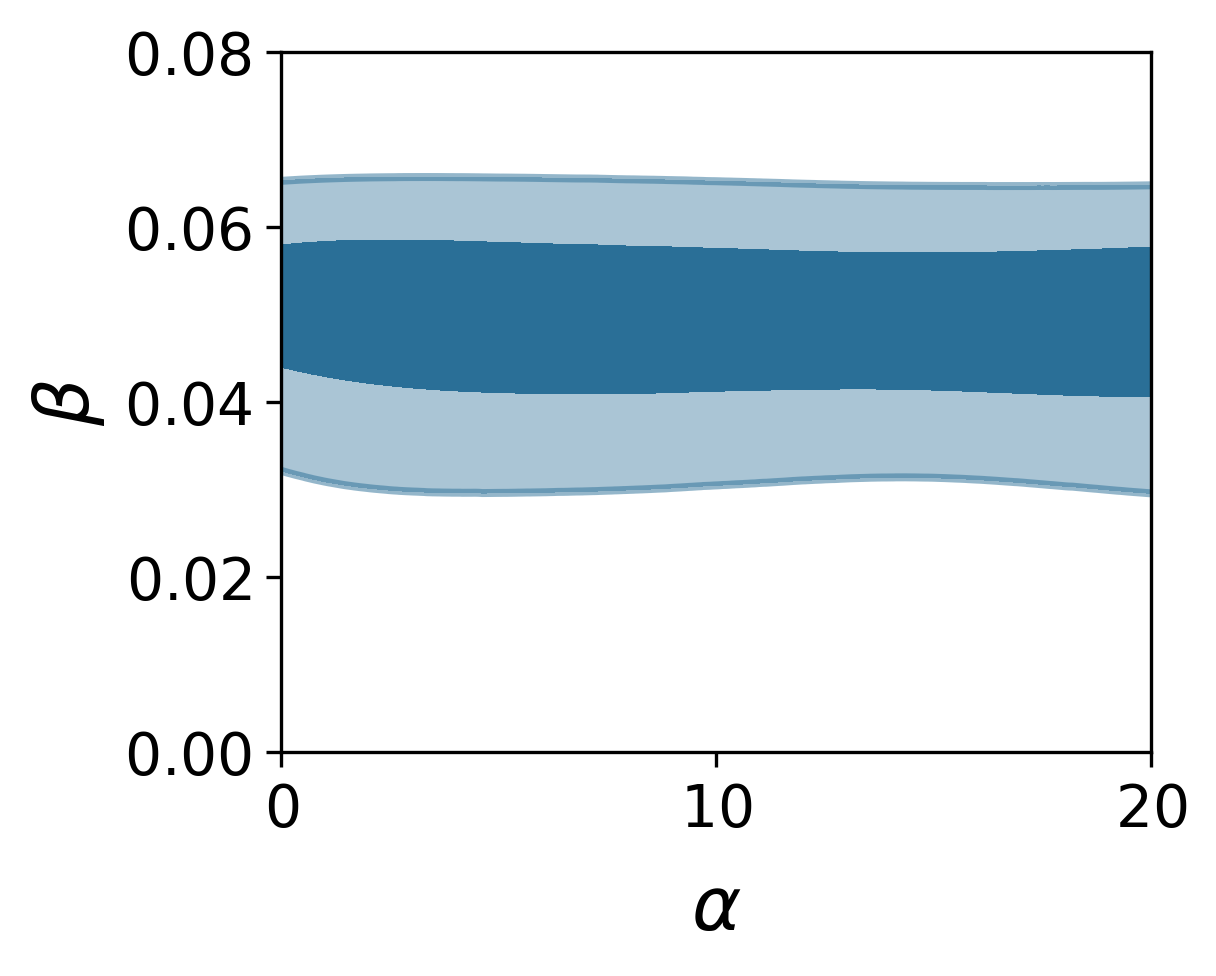}\\[-2pt]
\end{minipage}\hspace{0.06\textwidth}
\begin{minipage}{0.36\textwidth}
\centering
\includegraphics[width=\linewidth]{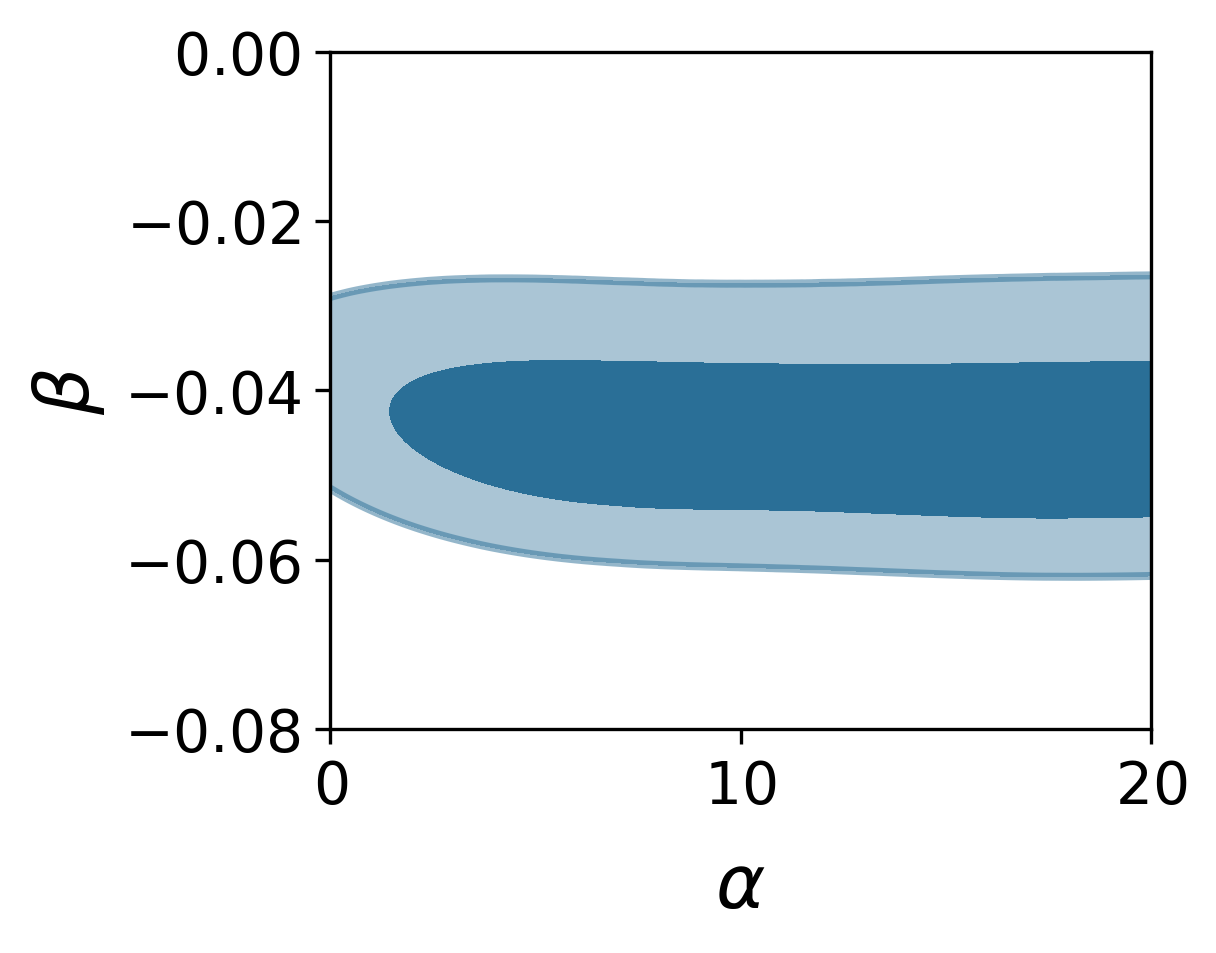}\\[-2pt]
\end{minipage}
\caption{ Posterior distributions for $(\alpha,\beta)$ in the minimum-start CQ-SUGRA branch. Left: $\beta>0$ and right: $\beta<0$. }
\label{fig4}
\end{figure}

The joint posterior distribution of $\alpha$ and $\beta$ is shown in the left panel of Fig.~\ref{fig4}, and the constraints on all free parameters are summarized in Tab.~\ref{tab2}.  We find that the values of $\Omega_m$ and $H_0$ are in good agreement with those in the minimally coupling case. While observations cannot provide an effective constraint on $\alpha$,  $\beta$ is constrained to be $\beta=0.049^{+0.010}_{-0.007}$ at $1\sigma$ CL. This indicates that a non-minimal coupling  between quintessence and CDM is favored at more than {$4\sigma$}, which is consistent with the findings from  the attractor-start branch.  

\begin{figure*}[t]
\centering
\begin{minipage}{0.48\textwidth}
\centering
\includegraphics[width=\linewidth]{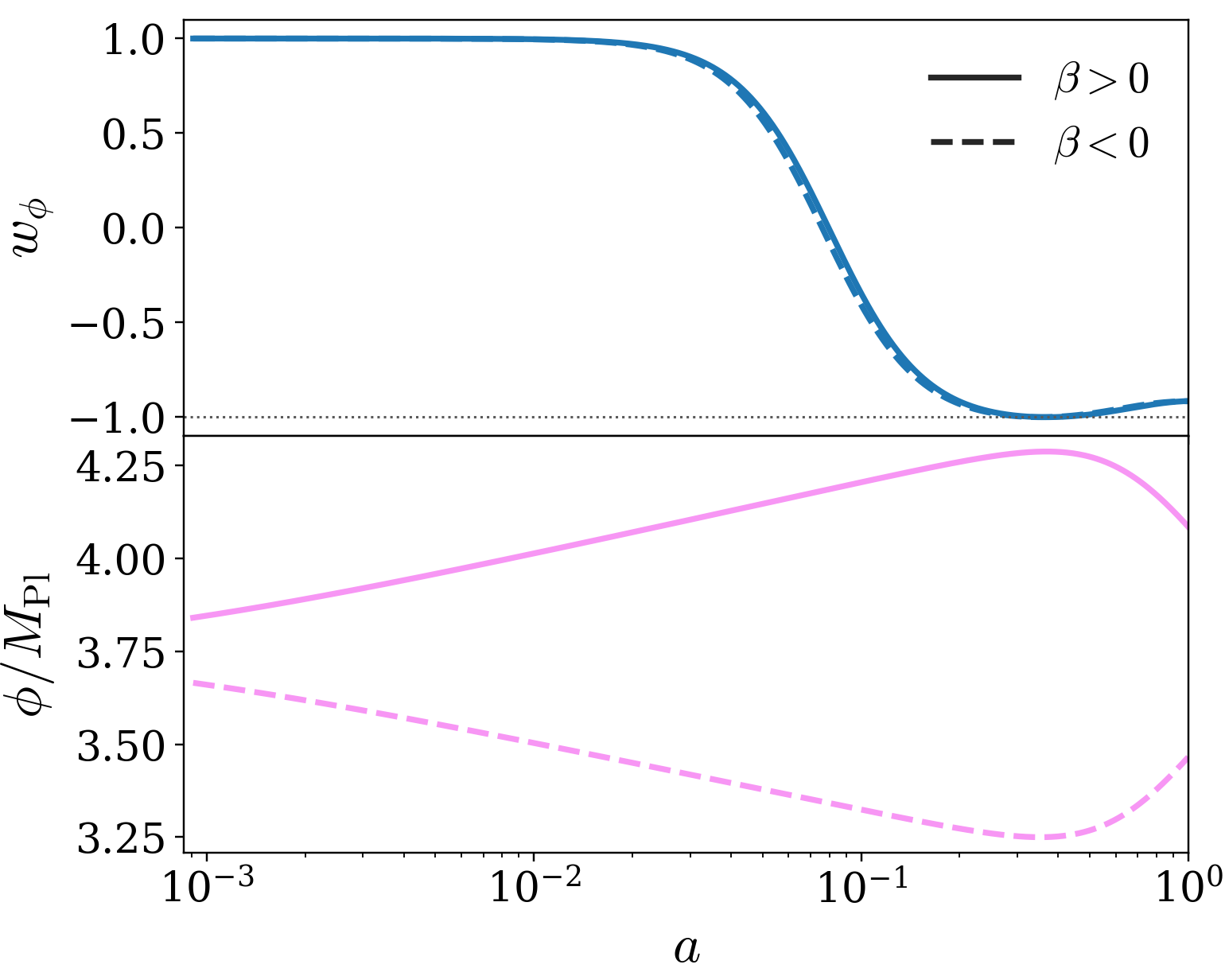}
\end{minipage}\hfill
\begin{minipage}{0.48\textwidth}
\centering
\includegraphics[width=\linewidth]{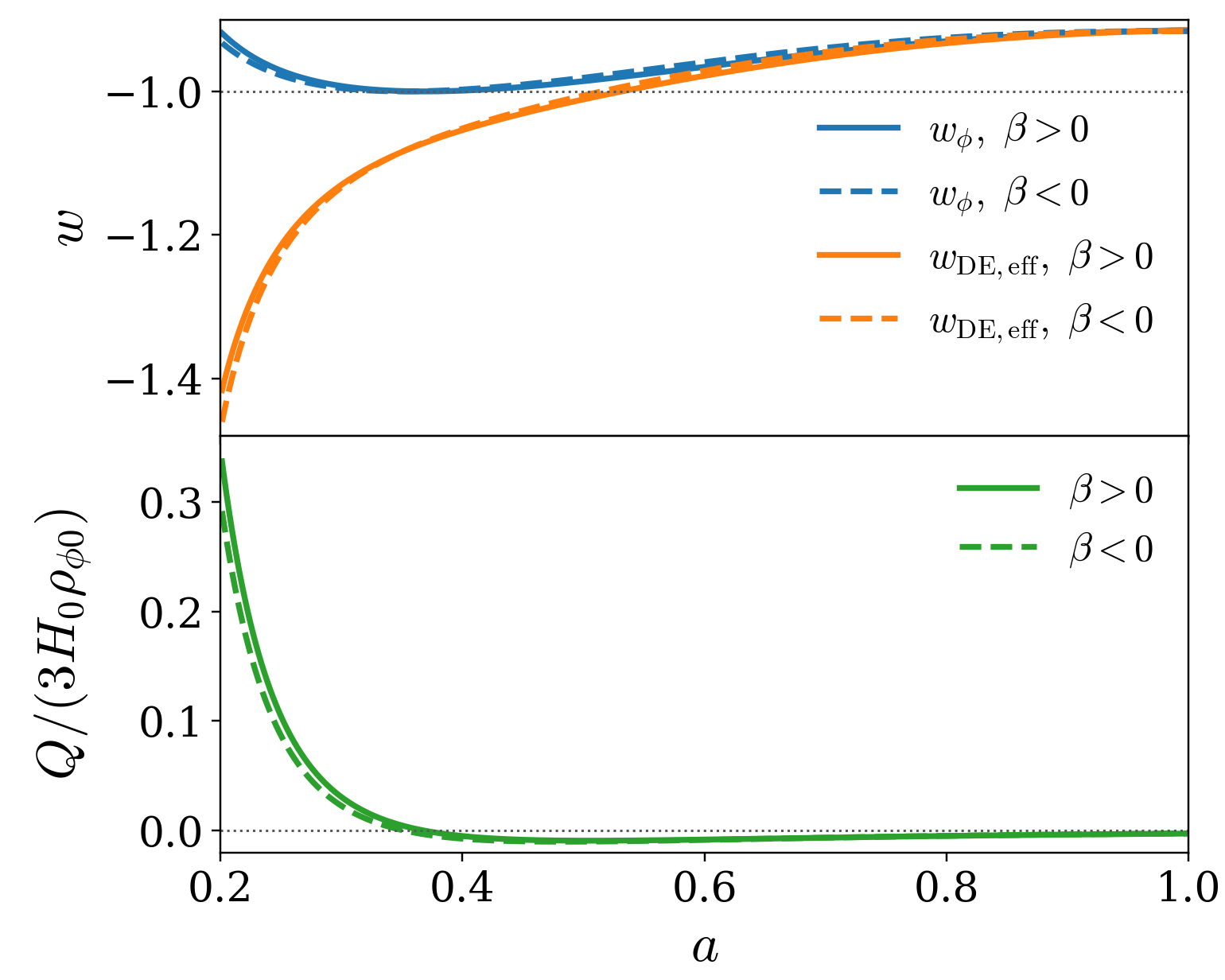}
\end{minipage}
\caption{Evolution of $\phi$, $w_\phi$, $w_{{\rm DE},{\rm eff}}$, and $Q$ for the minimum-start CQ-SUGRA branch. The solid lines denote the $\beta>0$ branch, while the dashed lines denote the $\beta<0$ branch.}
\label{fig5}
\end{figure*}

Fig.~\ref{fig5} displays the evolutions of $\phi$,   $w_\phi$,  $w_\mathrm{DE,eff}$  and $Q$. In contrast to the case of the attractor-start branch, $\phi$ increases initially with cosmic expansion and later decreases after  reaching its maximum. This means that $\dot{\phi}$ is positive initially,  becoming  negative with cosmic expansion, which indicates $Q>0$ before the turning point of the scalar field and $Q<0$ afterwards. This behavior is also depicted in the right panel of Fig.~\ref{fig5},  where the evolution of $Q$ is presented. Consequently, energy is initially transferred  from CDM to the scalar field, followed by a transfer from the scalar field to CDM.  Similar to the case of the attractor-start branch, in the early era, quintessence is dominated by its kinetic energy before evolving into  an era dominated by  potential energy. So, $w_\phi$ decreases initially with cosmic expansion. After the transition of $Q$ from positive to negative, $w_\phi$ increases with cosmic expansion, which contrasts with the behavior shown in Fig.~\ref{fig3}  where $w_\phi$ decreases with cosmic expansion. The value of $w_\phi$  is alway larger than $-1$. 
Moreover, the evolution of $w_{\mathrm{DE}, \mathrm{eff}}$  differs from $w_\phi$ at the late times. $w_{\mathrm{DE}, \mathrm{eff}}$ is below $-1$ before the bounce and crosses through the phantom divide at $a\simeq0.53$ ($z\simeq0.88$), after which it evolves to $w_{{\rm DE},{\rm eff}}>-1$. Therefore, the effective equation of state for  CQ-SUGRA can cross the $-1$ line from below  to above. This crossing behavior is favored by DESI BAO as indicated by  the value of   $\Delta\chi^2=-12.90$.

\begin{itemize}
\item $\beta<0$:
\end{itemize} 

The posterior distributions  for $ \beta$ and $\alpha$ are shown in the right panel of Fig.~\ref{fig4}. Unlike the case of  $\beta>0$, the data can give a lower limit on $\alpha$, i.e., $\alpha>4.57$ at $1\sigma$ CL, as summarized in Table~\ref{tab2}. The coupling is constrained to be $\beta=-0.045^{+0.006}_{-0.009}$, departing from zero significantly (more than 4$\sigma$). The preferred values of  $\Omega_m$ and $H_0$ are $\Omega_m=0.31\pm0.0041$ $H_0=67.43\pm0.35~{\rm km\,s^{-1}\,Mpc^{-1}}$, which are the maximum and minimum values  of $\Omega_m$ and $H_0$, respectively, shown in Table~\ref{tab2}.

From Fig.~\ref{fig5} where the evolutions of $\phi$,   $w_\phi$,  $w_\mathrm{DE,eff}$  and $Q$ are plotted, we find that  the main difference from the $\beta>0$ branch is the direction of the field motion. By the initial epoch, the interaction has displaced the field to the left side of the SUGRA minimum, such that $\phi<\sqrt{\alpha}M_{\rm Pl}$, and the field continues to move toward smaller $\phi$. It  turns around near $a\simeq0.35$ and rolls back toward larger $\phi$, with $w_\phi$ approaching $-1$ near the turning point. The evolutionary curves of $w_\phi$, $w_{{\rm DE},{\rm eff}}$ and $Q$ differ  only slightly from  those obtained in the $\beta>0$ case. This minor difference arises because,  for the case of $\beta<0$,  the roll speed of scalar field decreases more rapidly than that in the $\beta>0$ case, as the potential is slightly steeper for the former, which can be observed in Fig.~\ref{fig1}.  This effect causes  $w_{{\rm DE},{\rm eff}}$ to cross the $-1$ line, which appears at $a\simeq0.51$ ($z\simeq0.94$), to occur slightly earlier than in the case of $\beta>0$. This scenario appears to be favored by observations,  as indicated by $\Delta\chi^2=-13.09$, which is the minimum value shown in Tab.~\ref{tab2}.

\subsection{Discussions}

We have found that the  allowed values of $\Omega_m$ and $H_0$ in the SUGRA quintessence, both with and without coupling,  change negligibly and are consistent well those obtained in the $\Lambda$CDM model. Therefore, dynamical quintessence does not contribute to resolving the Hubble constant tension. Coupling between quintessence and CDM is strongly supported  by observations, as the value of $|\beta|$ departs from zero at more than {$4\sigma$}. Moreover, observations prefer the minimum-start CQ-SUGRA model,  as its values of $\chi^2_\mathrm{min}$ are  approximately 13  lower  than that of the $\Lambda$CDM model. This preference arises because the minimum-start CQ-SUGRA model exhibits a sign-change interaction between two dark sector. This interaction leads to a crossing of the effective equation of state for dark energy  from less than $-1$ to greater than $-1$ in the near past. Thus, dynamical dark energy with a phantom crossing of the equation of state is favored by observations.

\begin{figure*}[t]
\centering
\begin{minipage}{0.48\textwidth}
\centering
\includegraphics[width=\linewidth]{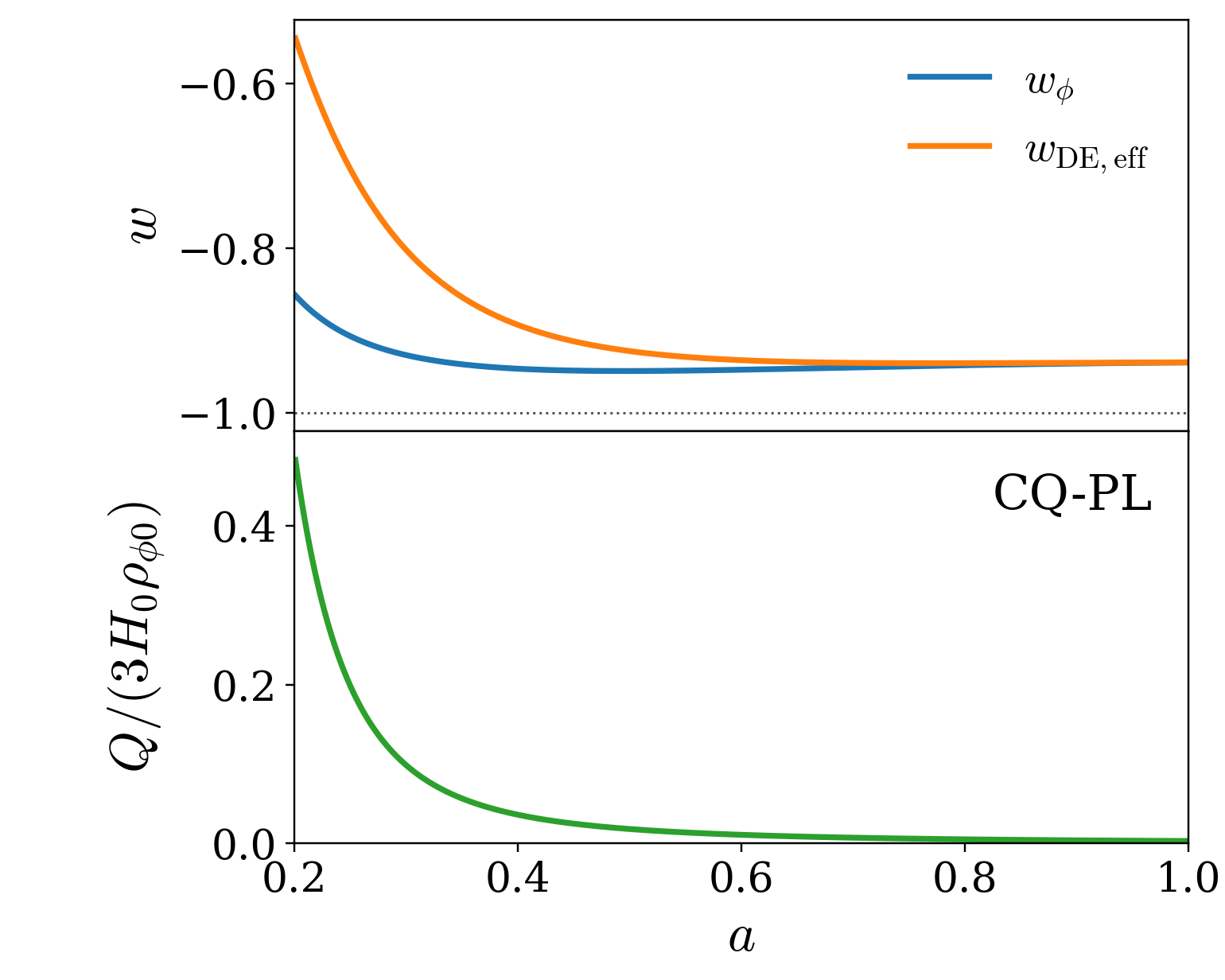}
\end{minipage}
\caption{Evolution of $w_\phi$, $w_{{\rm DE},{\rm eff}}$, and $Q$ for  the CQ-PL  model.}
\label{fig6}
\end{figure*}

 To compare the CQ-SUGRA  with other coupled quintessence models, we also consider the CQ-PL model. The results are summarized in Table.~\ref{tab3} and the evolutions  of  $w_\mathrm{DE, eff}$ and $Q$ are presented in Fig.~\ref{fig6}. Similar to the case of CQ-SUGRA, the values of $\Omega_m$ and $H_0$ align with those  given in the $\Lambda$CDM model, and coupling is strongly preferred, as $\beta$ deviates from zero at more than {$4\sigma$}.  However, the values of $\Delta\chi^2$ in the minimum-start CQ-SUGRA   are much smaller than the one in the CQ-PL. This difference arises  because  $w_\mathrm{DE,eff}$ can not across the $-1$ line in the CQ-PL model,  due to the energy transfer  not changing sign, as shown in Fig.~\ref{fig6}. 

Table~\ref{tab3} also summarizes the constraints on the $w_0w_a$CDM model.  For this model, the mean values of $\Omega_m$ and $H_0$ are larger and smaller, respectively,  than those of all the other models we considered. The value of $\Delta\chi^2$ in the  $w_0w_a$CDM model is slightly lower than that in the minimum-start CQ-SUGRA model, but this difference is only about $1.5$, which is insufficient to determine which model is favored by observations.

\begin{table*}[t]
\centering
%\color{red}
\caption{Summary of the  constraints on the $w_0w_a$CDM and CQ-PL  models.}
\label{tab3}
\small
\setlength{\tabcolsep}{9pt}
\begin{tabular}{lcc}
\hline\hline
Parameter & $w_0w_a$CDM & CQ-PL \\
\hline
$w_0$ & $-0.811\pm 0.055$ & $-$ \\
$w_a$ & $-0.69^{+0.23}_{-0.20}$ & $-$ \\
$\alpha$ & $-$ & $0.46\pm 0.18$ \\
$\beta$ & $-$ & $0.051^{+0.011}_{-0.008}$ \\
$H_0$ & $67.31\pm 0.54$ & $67.82\pm 0.53$ \\
%\hline\hline
$\Omega_m$ & $0.3132\pm 0.0053$ & $0.2996\pm 0.0049$ \\
%$S_8$ & $0.8244\pm 0.0087$ & $0.8192\pm 0.0087$ \\
\hline\hline
$\Delta \chi^2$ & $-14.48$ & $-9.83$ \\
\hline\hline
\end{tabular}
\end{table*}

\section{Conclusions}
\label{sec4}

Quintessence with a SUGRA potential has the distinctive ability to realize dynamical dark energy while also admitting a cosmological-constant-like behavior when the scalar field settles near the minimum of the potential. Moreover, when SUGRA quintessence is coupled to cold dark matter, the effective equation of state of dark energy can cross the phantom divide. Motivated by these features, we have tested both coupled and uncoupled SUGRA quintessence models using DESI BAO, DES-Dovekie SNIa, and Planck CMB data.

We find that neither the coupled nor the uncoupled SUGRA quintessence model significantly increases the inferred value of $H_0$, and therefore the Hubble tension remains unresolved in this framework. However, the data strongly favor a coupling between SUGRA quintessence and cold dark matter, with the coupling parameter deviating from zero at more than $4\sigma$. In particular, the minimum-start coupled SUGRA quintessence model can generate a sign-changing interaction between dark energy and dark matter, which in turn drives the effective dark-energy equation of state across the phantom divide. This branch is favored by the data, yielding a value of $\chi^2_\mathrm{min}$ approximately 13 lower than that of $\Lambda$CDM. This result indicates that current observations favor an evolving dark-energy equation of state capable of crossing $w=-1$.

Interestingly, the difference in $\chi^2_\mathrm{min}$ between the minimum-start coupled SUGRA quintessence model and the phenomenological $w_0w_a$CDM model is very small, making it difficult to distinguish between them statistically with current data. The minimum-start coupled SUGRA model therefore provides a physically motivated, field-theoretic realization of the evolving dark energy behavior suggested by recent observations, while achieving a fit comparable to that of the CPL parametrization.

Finally, we note that, during the preparation of this work, it was shown that a sign change in the energy transfer between dark matter and dark energy can also occur in coupled quintessence with an exponential potential, for a coupling of the form $1+\frac{\beta}{2}\phi^2$~\cite{wang2026signswitch}. This suggests that sign-changing interactions may be a broader feature of coupled scalar-field dark-energy models and deserve further investigation.

\begin{acknowledgments}
This work was supported in part by the NSFC under Grant  Nos. 12275080 and 12075084,  and the Innovative Research Group of Hunan Province under Grant No.~2024JJ1006.

\end{acknowledgments}

\end{document}